\documentclass{Rinton-P9x6}

\def\be {\begin{equation}}
\def\ee {\end{equation}}
\def\ba {\begin{eqnarray}}
\def\ea {\end{eqnarray}}
\def\nn {\nonumber}
\def\no {\noindent}
\def\D  {\Delta}
\def\e  {\epsilon}
\def\la {\label}
\def\le {\left}
\def\ri {\right}
\def\pa {\partial}
\def\f {\frac}

\def\no {\noindent}
\def\bi {\begin{itemize}}
\def\ei {\end{itemize}}

\def\ra {\rangle}
\def\vs {\vspace}

\begin{document}

\title{Rapid Data Search using Adiabatic Quantum Computation}

\author{Daria Ahrensmeier$^\sharp$, Saurya Das$^\flat$
\footnote{Primary Author}~~~, Randy Kobes$^\sharp$, \\ 
Gabor Kunstatter$^\sharp$, Haitham Zaraket$^\sharp$}

\address{$^\sharp$Dept. of Physics and The Winnipeg Institute for Theoretical
Physics\\ The University of Winnipeg, \\
515 Portage Avenue, Winnipeg, MB - R3B 2E9, CANADA\\
E-mail: dahrens,randy,gabor,zaraket@theory.uwinnipeg.ca}


\address{$^\flat$Dept. of Mathematics and Statistics, \\
University of New Brunswick, Fredericton, NB - E3B 5A3, CANADA
\\E-mail: saurya@math.unb.ca}


\maketitle

\abstracts{
We show that by a suitable choice of time-dependent Hamiltonian,
the search for a marked item in an unstructured database can be
achieved in unit time, using Adiabatic Quantum Computation. 
This is a considerable
improvement over the ${\cal O}(\sqrt{N})$ time required in previous
algorithms. The trade-off
is that in the intermediate stages of the computation process, the ground
state energy of the computer increases to a maximum of ${\cal O}(\sqrt{N})$,
before returning to zero at the end of the process.
}

\section{Adiabatic Quantum Computation} 

Adiabatic Quantum Computation (AQC) is a new paradigm in quantum 
computation, in which an initial state $|\Psi_0\ra$ is {\it adiabatically}
transformed into a final state $|\Psi_1\ra$ by means of a time-dependent
Hamiltonian \cite{adia} :
\ba
H(s) &=& f(s) H_0 + g(s) H_1 \la{ham1} \\
\mbox{where}~~~H_0 &=& I - |\Psi_0\ra\langle\Psi_0|  \la{ham2} \\
H_1 &=& I - |\Psi_1\ra\langle\Psi_1| ~~\la{ham3}   
\ea
($s(t)$ is a time-like parameter, which monotonically 
increases with time $t$, such that $s(0)=0$ and $s(T)=1$, where $T$ is
the required running time of the AQC). 
The boundary conditions on $f(s)$ and $g(s)$ are as follows:
\be
f(0) = g(1) = 1~~~,~~~f(1)=g(0)=0~~.
\la{bc1}
\ee
Note that $|\Psi_0\ra$ and $|\Psi_1\ra$ are ground states of 
$H_0$ and $H_1$, respectively. The {\it adiabaticity condition},
which must be preserved at all times, is given by:
\be
\f{\le| \langle -|(dH/dt)|+\ra \ri|}{(E_+-E_-)^2} \leq \e ~~,
\ee
$$\mbox{where}
~~|\pm \ra = \mbox{ground state \& first excited state}~~:
~~H|\pm\ra = E_\pm |\pm\ra \nonumber~~.
$$
The problem is to estimate the running time $T$ for a 
given computational problem implemented via an AQC.

\section{Data Search Problem} 

For a completely unstructured database of  
$N$ items, one would like to find a marked item
(say $``m''$) in the shortest possible time. Schematically:

\vs{.2cm}
$\underbrace{|\cdot|\cdot|\cdot|
\cdots \overbrace{|m|}^{\mbox{Marked item}} 
\cdots |\cdot|\cdot|\cdot|}_
{\mbox{unstructured 
database of}~N~\mbox{items}}$
~~$\rightarrow$~Find $m$ in shortest possible time

\vs{.1cm}
\no
Now, it is well-known that 
classically, ${\cal O}(N)$ steps are required on an average to find $m$.
On the other hand, it was shown by Grover, that quantum mechanics
can be used to quadratically speed-up the process \cite{grover}. 
That is, by 
associating each item in the database with an eigenvector in a
$N$-dimensional Hilbert space (such that $m$ corresponds to the 
ket $|m\ra$), then starting with the symmetric superposition of
all states, and applying a series of unitary operations, one can 
evolve to the marked state in ${\cal O}(\sqrt{N})$ steps.

\section{Data Search Using AQC}

In this case, the adiabatic Hamiltonian (\ref{ham1}) 
replaces the set of unitary transformations mentioned in the 
previous section. The final state $|\Psi_1\ra$
is again the marked state $|m\ra$. 
Now, it can be proved, by closely following the arguments of \cite{fg,rc}, 
that the following inequality holds (see Appendix):
\be
\int_0^T g(s(t)) \geq \f{k\sqrt{N}}{4}~~,
\la{thm1}
\ee
(where $k$ is a constant of order unity). Note that previously 
\cite{rc} it had been assumed that
$g(s(t))=1-s =1-f(s(t)) \leq 1~~\forall s$, 
for which the above theorem implies that the running time
$ T \geq k\sqrt{N}/4$
which is at par with Grover's algorithm.
Result (\ref{thm1}) can be thought of 
as AQC generalization of the lower bound on the number of steps of any
search algorithm that has been proved previously \cite{bbbv}. 

The `gap' between the ground state and first excited state of the
Hamiltonian is given by \cite{dkk}:
\be
\D (s) =  \sqrt{(f-g)^2 +\f{4}{N}fg}
\la{gap1}
\ee
For the choice $g=1-f=1-s$, $\mbox{Min}(\D) = \D(1/2) = 1/\sqrt{N}$,
which is in conformity with the general idea that the running time
and minimum gap are inverses of each other. 
However, (\ref{thm1}) and (\ref{gap1}) suggest that by suitably 
changing $g(s(t))$, so as to increase the gap considerably,
the running time can be significantly reduced. 
We make such a choice, which satisfies the boundary conditions 
(\ref{bc1}):
\be
f = 1-s +\sqrt{N} s(1-s)~~~,~~~~
g = s +\sqrt{N} s(1-s)~.
\ee
It is easy to see that for the above choice, 
$\mbox{Min} (\D) = {\cal O}(1)$ and the running time is \cite{dkk}
$$ T = \f{1}{\e} \le( 1 + \f{\pi}{2} \ri) $$
which is a constant. The original gap and the modified gap are plotted in
figure 1 for $N=10,000$.

\begin{figure}[t]
~~~~~~~~~~~~~~~~\epsfxsize=6pc 
\epsfbox{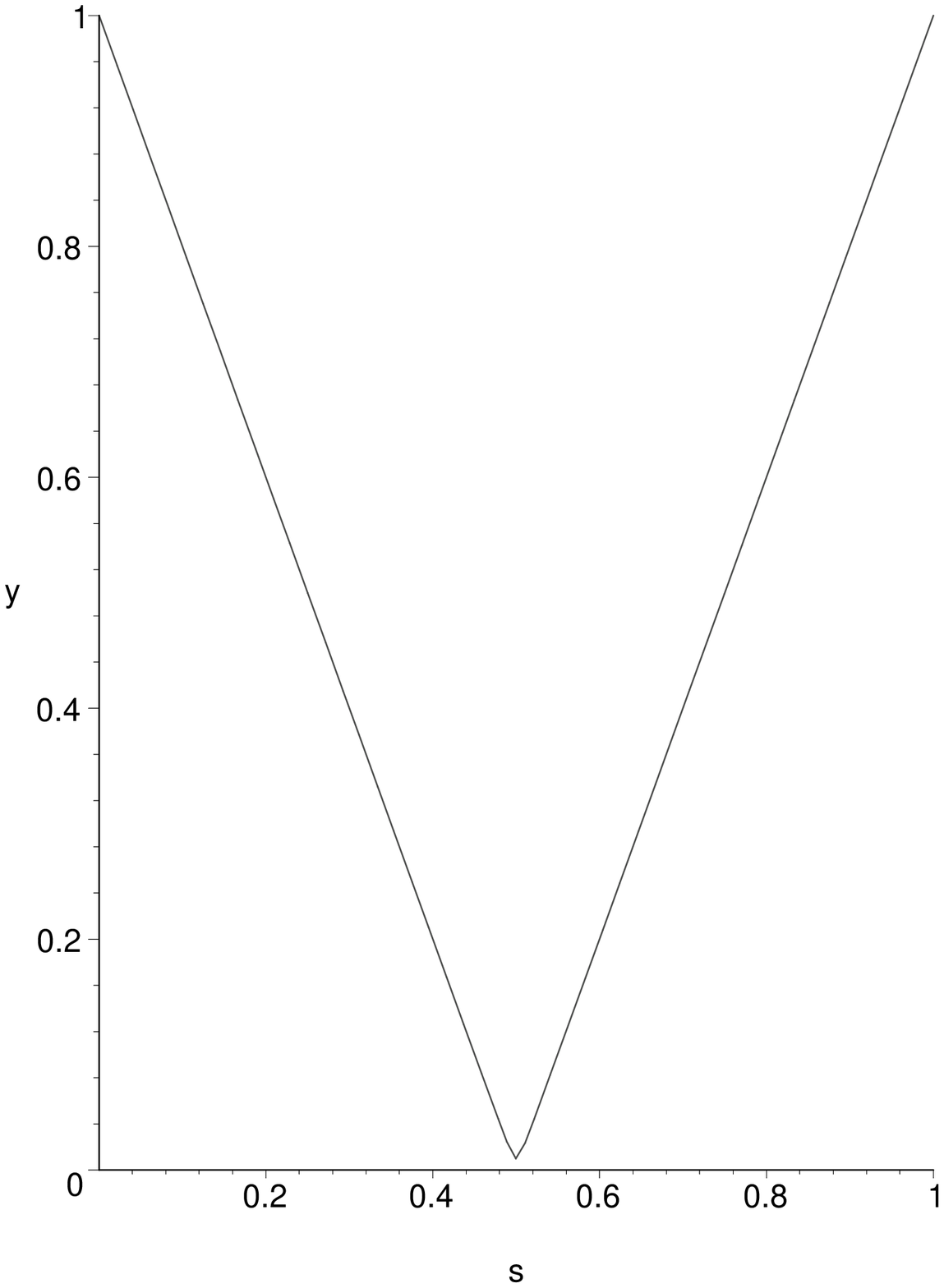} 
~~~~~~~~~~~~~~~~~~~~~\epsfxsize=6pc 
\epsfbox{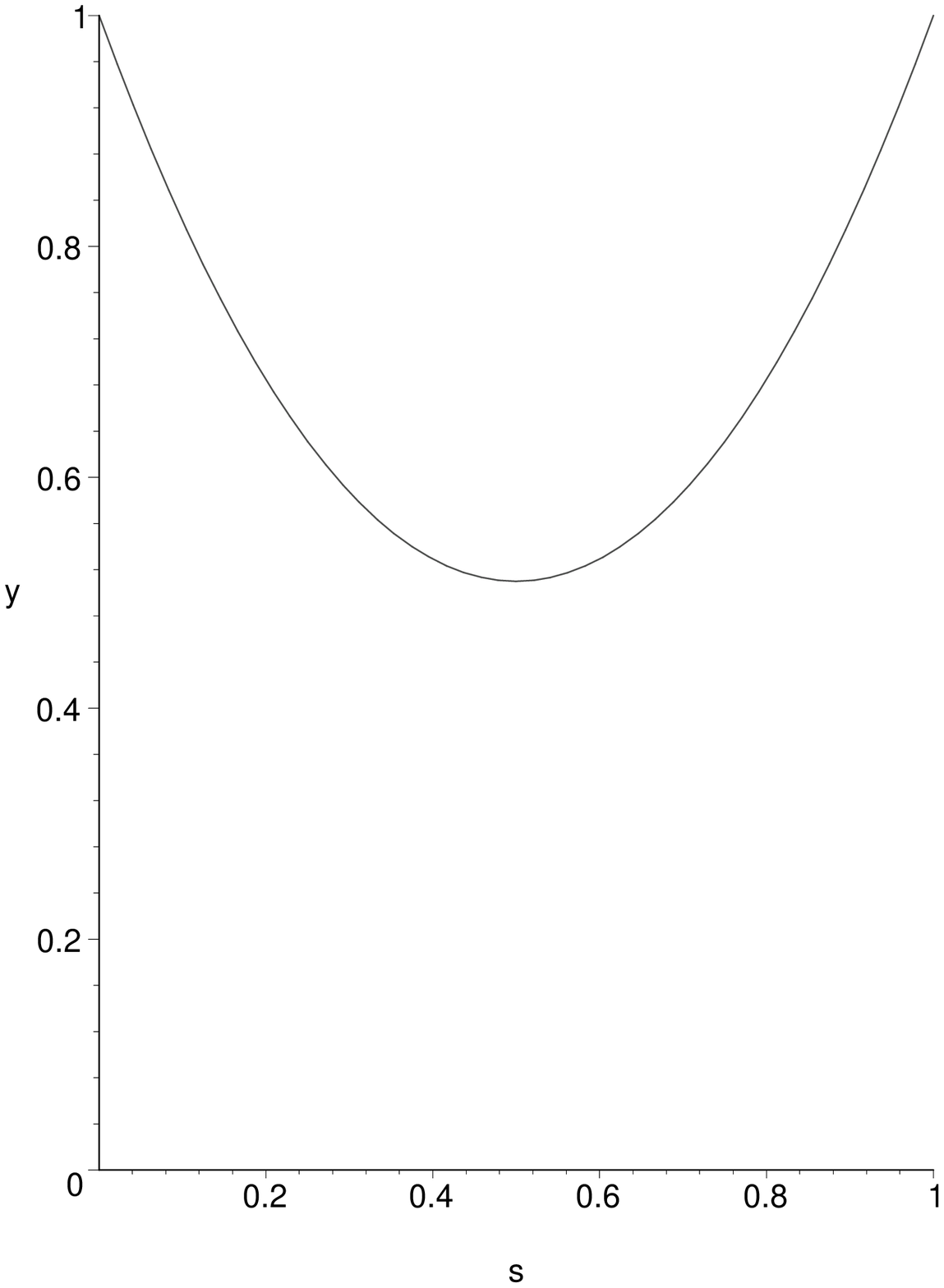} 
\caption{Original gap and modified gap $\D$ is plotted as a function of
$s: 0 \rightarrow 1$ for $N=10,0000$
\label{fig:gap1}}
\end{figure}

\section{Discussion}

\no
(1) Note that
although the ground state energy rises to ${\cal O}(\sqrt{N})$
for intermediate times, no energy is actually being used in the 
process. Thus, energy need not be regarded as a resource in this
context. Furthermore, by subtracting $E_-$ from
Hamiltonian (\ref{ham1}) at all times, the resultant 
ground state energy can be made to vanish,
without modifying the gap or the running time.\\ 
Error correcting processes would presumably entail heat losses,
commensurate with the second law of thermodynamics. \\
(3) A larger gap in our case may signal greater fault-tolerance than
previously considered Hamiltonians \cite{robust}. \\ 
(4) Work is in progress towards similar considerations for structured
data searches, to see whether similar speed-ups are possible
\cite{dkk}.

\section*{Acknowledgments}
The authors would like to thank E. Farhi and S. Gutmann for 
useful discussions. 
S.D. thanks V. Husain for useful comments. This work was 
supported by the Natural Sciences and Engineering Research Council
of Canada.

\section*{Appendix}

To prove (\ref{thm1}), first write (\ref{ham1}) as \cite{fg,rc}:
\be
H (s) = H_1 (s) + H_{2m} (s) ~~,
\ee
where
\ba
H_1 (s) &=& \le( f(s) + g(s) \ri) - f(s) |\psi_0\ra
\langle \psi_0|  \la{h3} \\
H_{2m} (s) &=& - g(s) |m\ra \langle m |  ~~.
\ea
Consider two computers $|\psi_m,t\ra$ and
$|\psi_{m'},t\ra$ respectively at any instant $t$, evolving to states
$|m\ra$ and $|m'\ra$. 
The Schr\"odinger equations are:
\be
i  \f{\pa}{\pa t} |\psi_{m,m'},t\ra = \le( H_1 +
H_{2m,2m'} \ri) |\psi_{m,m'},t\ra \la{sch1} \\
%
\ee
subject to the boundary conditions:
\be
|\psi_m, 0 \ra = |\psi_{m'}, 0 \ra = |\psi_0 \ra~;  
~~|\psi_m, T \ra = |m\ra ~,~ 
|\psi_{m'}, T \ra = |m' \ra~~~. \la{bc2}
\ee
{}From (\ref{sch1}), it follows that:
\be
{}  \f{\pa}{\pa t} \sum_{m,m'} \le[  1 - |\langle
\psi_m,t | \psi_{m'}, t \ra^2 \ri]  \nn 
\leq  {4 N^{3/2}}{}  g(s)~~.  \la{sum1}
\ee
Integrating (\ref{sum1}) from $t=0$ to $t=T$, and
using the boundary conditions 
(\ref{bc2}), we get:
\be
\sum_{m,m'} \le[ 1 - |\langle \psi_m,T|\psi_{m'},T \ra
|^2 \ri] 
\leq {4 N^{3/2}}{}  \int_0^T g(s(t)) dt~~.
\la{time5}
\ee
Finally, using the fact: 
\be
1 - |\langle \psi_m,T|\psi_{m'},T\ra|^2 \geq
k~~~,\forall m \neq m'
\ee
which simply means that different computers evolve the
same initial
state to sufficiently different final states. This
yields (\ref{thm1}), for $N \gg 1$.

\end{document}